\documentclass[journal]{IEEEtran}

\usepackage{graphicx}
\usepackage{latexsym}
\usepackage{amsfonts}
\usepackage{amssymb}
\usepackage{amsbsy}
\usepackage{amsmath}
\usepackage{multicol}
\usepackage{float}
\usepackage{algorithm}
\usepackage[dvips]{epsfig}

\usepackage{makecell}
\usepackage{subcaption}
\usepackage{hyperref}
    
\usepackage{paralist}
\usepackage{caption}
\usepackage{fancyhdr}
\usepackage{epsfig}
\usepackage{threeparttable}
\usepackage{epsf,epsfig}
\usepackage{amsthm}
\usepackage[noadjust]{cite}
\usepackage{dsfont}
\usepackage{color}
\usepackage{enumerate}
\usepackage{setspace}
\usepackage{bbm}

\usepackage{array}
\makeatletter
\newcommand{\thickhline}{%
    \noalign {\ifnum 0=`}\fi \hrule height 1pt
    \futurelet \reserved@a \@xhline
}
\newcolumntype{"}{@{\hskip\tabcolsep\vrule width 1pt\hskip\tabcolsep}}
\makeatother

\usepackage{amsmath}
\usepackage{algorithm}
\usepackage[noend]{algpseudocode}
\algrenewcommand{\algorithmiccomment}[1]{\hskip3em$\triangleright$ #1}

\makeatletter
\def\BState{\State\hskip-\ALG@thistlm}
\makeatother

\def\SIR{\mathsf{SIR}}

\def\({\left(}
\def\){\right)}
\def\[{\left[}
\def\]{\right]}

\usepackage{dblfloatfix} 

\title{Opportunism in Dynamic Spectrum Access for 5G: \\A Concept and Its Application to Duplexing}
\begin{document}


\author{Jeemin~Kim,~\IEEEmembership{Student~Member,~IEEE,} 
Soo-Min~Kim,~\IEEEmembership{Student~Member,~IEEE,}
Han~Cha,~\IEEEmembership{Student~Member,~IEEE,}
Jinho~Choi,~\IEEEmembership{Student~Member,~IEEE,}
Seung-Woo~Ko,~\IEEEmembership{Member,~IEEE,}\\
Chan-Byoung Chae,~\IEEEmembership{Senior~Member,~IEEE,}
and~Seong-Lyun~Kim,~\IEEEmembership{Senior~Member,~IEEE}
\thanks{J. Kim, S.-M. Kim, H. Cha, J. Choi, and S.-L. Kim are with  Yonsei University, Korea (email: \{jmkim, chan, jhchoi\}@ramo.yonsei.ac.kr, \{sm.kim, cbchae, slkim\}@yonsei.ac.kr). S.-W. Ko is with The University of Hong Kong, Hong Kong, (email: swko@eee.hku.hk). J. Kim and S.-M. Kim are co-1st authors.}
}

\maketitle

\begin{abstract}
With the envisioned massive Internet-of-Things (IoT) era, one of the challenges for 5G wireless systems will be handling the unprecedented spectrum crunch.
A potential solution has emerged in the form of spectrum sharing, which deviates from a monopolistic spectrum usage system.
This article investigates the medium access control (MAC) as a means of increasing the viability of the spectrum sharing technique.
We first quantify the opportunity of spectrum access in a probabilistic manner, a method referred to as \emph{opportunity probability} (OP).
Based on the OP framework, we propose a random MAC algorithm in which the access of a node is randomly determined with its own OP value.
As a possible application of our OP based random MAC, we propose a hybrid half duplex (HD)/full duplex (FD) communication where each pair decides the duplexing mode according to the OP values of the two pair nodes.
This approach fits well with the spectrum sharing system since it enables a flexible operation for the spectrum access according to the spectrum usage level.
From the numerical analysis, we validate the feasibility and verify the performance enhancements by implementing an FPGA based real-time prototype. Measurements and numerical results confirm that the proposed architecture can achieve up to 4 times higher system throughput than conventional LTE-TDD (time division duplex) systems.
\end{abstract}


\IEEEpeerreviewmaketitle

\section{Introduction}

\IEEEPARstart{T}{he} way people live and get around has been altered by the emergence of the sharing economy (e.g. car or house sharing).
This form of sharing improves the efficiency of resource utilization.
It breaks through the conventional ownership-based usage system where resources are often used sub-optimally.
The effect of sharing increases when the resource is scarce and more valuable, such as a spectrum resource in wireless networks.
The value of spectrum, especially in 5G systems aiming to accommodate $1,000$-fold more \emph{Internet-of-Things} (IoT) devices, will skyrocket.
The traditional monopolistic spectrum allocation will be unable to cope with such a rise in demand.
Hence, more and more researchers are turning their interest to spectrum sharing in which multiple networks coexist in one spectrum band \cite{qualcomm}.


This article considers how to design a spectrum sharing network that can address the unevenly utilized spectrum in time and space.
Recent measurements \cite{sanjit15} show that several frequency bands are very much under-utilized, while other bands are over-utilized.
This imbalance can be solved through spectrum sharing, which permits the access of under-utilized spectrums.
To this end, the sharing network should accurately track the degree of spectrum usage level in real time.

In the conventional spectrum sharing technique such as carrier-sense multiple access (CSMA) or listen-before-talk (LBT) \cite{huang11}, the network utilizes spectrum sensing to check the spectrum usage level.
This way, the network deterministically decides whether the access opportunity exists or not.
Drawbacks to this technique, however, include the usage of spectrum varying spatiotemporally and the deterministic decision scheme.
Firstly, since each sensor only provides the local sensing information, uncertainty arises regarding the detection of spectrum usage.
A straightforward solution can be made by measuring the spectrum occupancy with an ultra-dense resolution, but this requires an expensive installation.
Second, the deterministic opportunity decision may incur frequent transmission collisions because the access opportunity is spatially correlated.
Nodes having access opportunity are likely to be located close together.
In this case, the deterministic access decision may make neighboring nodes simultaneously transmit, thereby causing the collision and energy waste.

\begin{figure*}
\centering 
{\includegraphics[width=15cm]{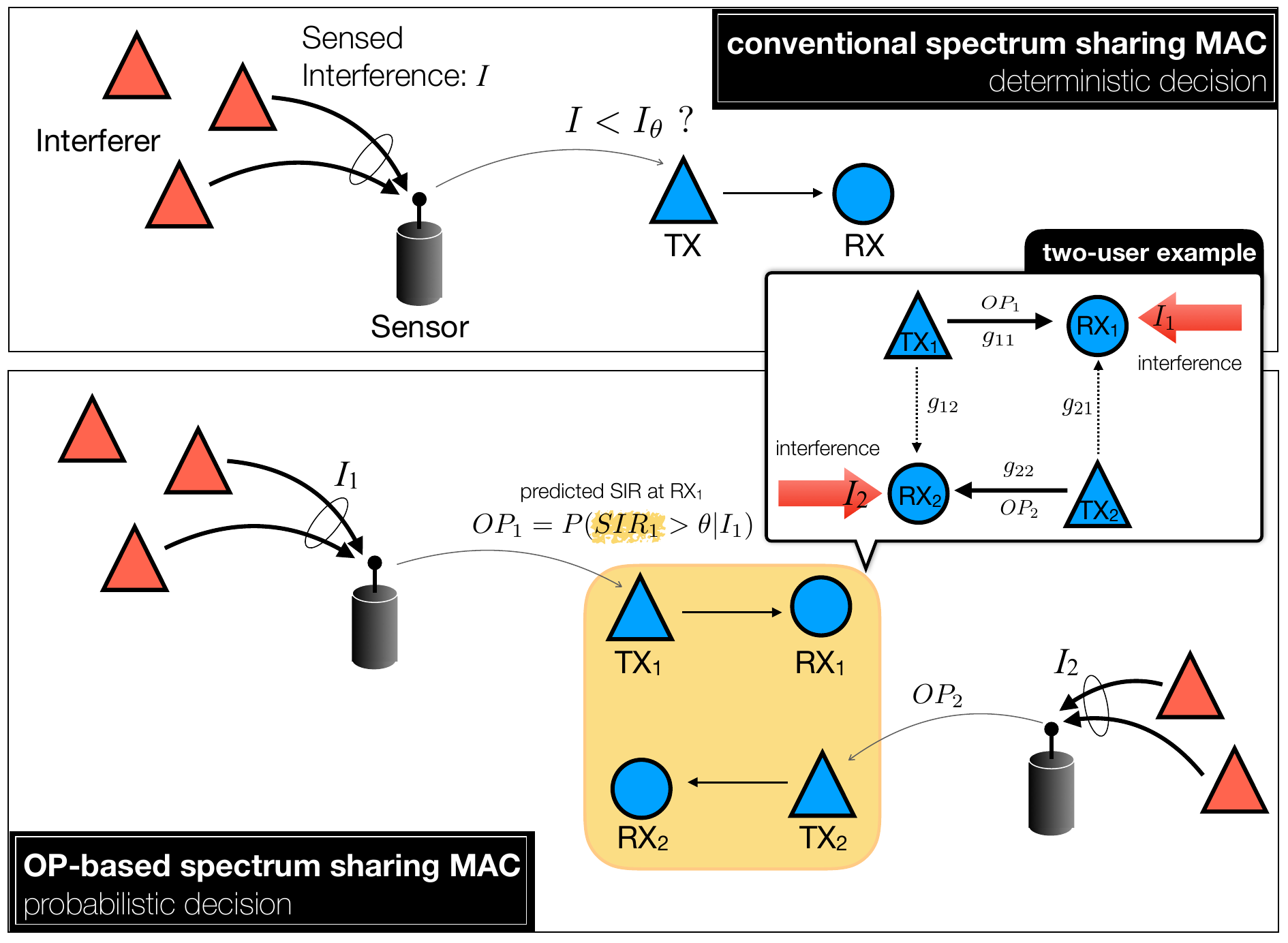}}
\caption{A spectrum sharing system: conventional deterministic MAC and OP-based MAC. With the deterministic MAC, each transmitter (TX) decides to transmit when the sensed interference $I$ is smaller than a threshold $I_\theta$. With the OP-based MAC, each TX randomly transmits with the OP, the transmission success probability.
The value $g_{ij}$ is the channel gain between TX~$i$ and RX~$j$.\label{Fig:system}} 
\end{figure*}

We thus propose a new sharing paradigm in which the spatiotemporal randomness of the spectrum usage level is quantified in a probabilistic manner so that with this probability a decision is made regarding the access of the node.
Specifically, we predict the spectrum usage level at locations where sensors are not installed and represent the access opportunity level as a probability, namely opportunity probability (OP).
The probabilistic approach enables one to check the opportunity using only a small number of sensors, achieving a cost efficient network, a priority for the 5G system \cite{ericsson}.
Each node randomly decides whether to access the spectrum or not with a probability according to the OP value.
The proposed OP based medium access control (MAC) scheme can resolve the transmission collision problem by decreasing the concurrent transmission in an area where high OP exists, while guaranteeing the energy efficiency.
Further, in an area with a low OP value, our OP based MAC scheme allows a few transmissions of nodes whereas no node may transmit with the deterministic MAC scheme.
It thereby increases the spatial reuse.

Such an OP framework can widely be utilized for the spectrum sharing network operation.
As a possible application, this article utilizes the OP based MAC to decide the duplex mode of pair nodes that are capable of full duplex (FD) communication, a technique in which the paired two nodes can simultaneously transmit on the same frequency band.
In the spectrum sharing network, for a harmonious coexistence among networks, each pair should decide the duplex mode among FD, half duplex (HD), and no transmission according to the OP value.
For instance, a pair node can transmit in FD mode only if the two nodes both have high OP values.
In this light, we suggest a hybrid HD/FD system in which each pair node randomly decides to access the spectrum with a given probability, OP, and thereby the duplex mode of its pair can be determined.
By so doing, the probability that a pair utilizes the FD mode naturally increases as the pair has a high OP value.
In the rest of this article, we report on the optimal access probability according to the OP value, considering the characteristic of IoT networks.

\section{Preliminaries}
This section briefly visits the basic concepts of our OP based MAC design and its application to the hybrid HD/FD communication.

\subsection{Spectrum Sharing Network Architecture}
Consider a general spectrum sharing system composed communication nodes, spectrum sensors and a centralized server (see Fig.~\ref{Fig:system}).
Spectrum sensors periodically measure the interference level at their locations to check the spectrum usage level.
The measured results are sent to a centralized server.
The sever calculates an OP value at every location with a given measured interference level of the nearest sensor.

As a simple example, we consider a two-user case as depicted in Fig.~\ref{Fig:system}.
Each pair has a respective OP value given from the server.
Based on this OP each transmitting node makes a decision for transmission, which affects not only the pair's communication performance but also the other pair's link quality.
Therefore, the access probability should be determined by regarding both, which will be more discussed further in Section II-C.


\begin{figure*}
	\centering 
	{\includegraphics[width=16cm]{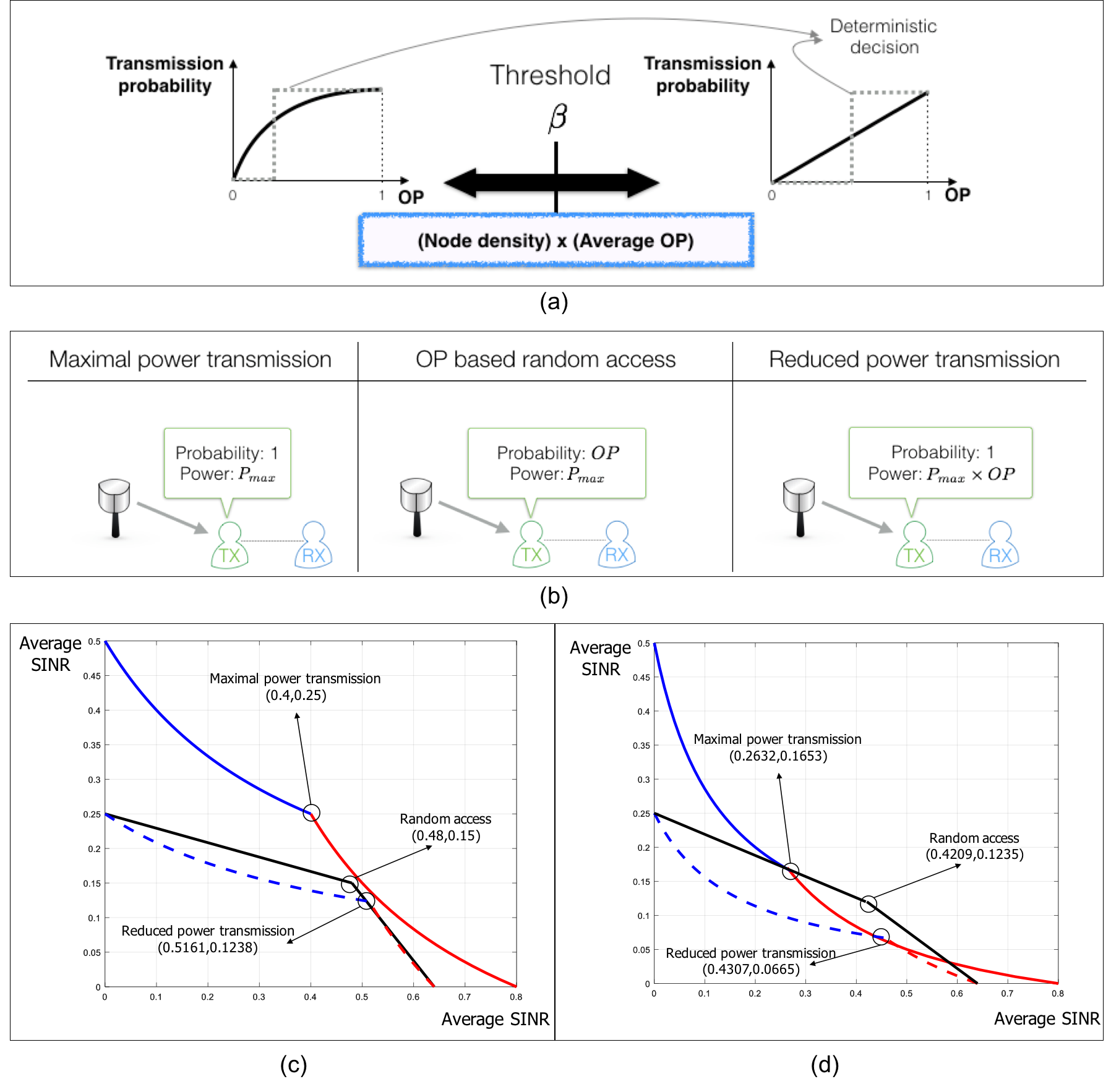}}
	\caption{a) Proper shapes of function for controlling the transmission probability; b) Probability and power settings for each access mode; c) SINR region for the low level of mutual interference ($g_{11}=g_{22}=0.04, g_{12}=g_{21}=0.05, I_1=0.0125, I_2=0.02$); d) SINR region for the high level of mutual interference ($g_{11}=g_{22}=0.04, g_{12}=g_{21}=0.15, I_1=0.0125, I_2=0.02$).}\label{Fig:ProbvsPower}
\end{figure*}

\begin{figure*}
\centering 
{\includegraphics[width=17cm]{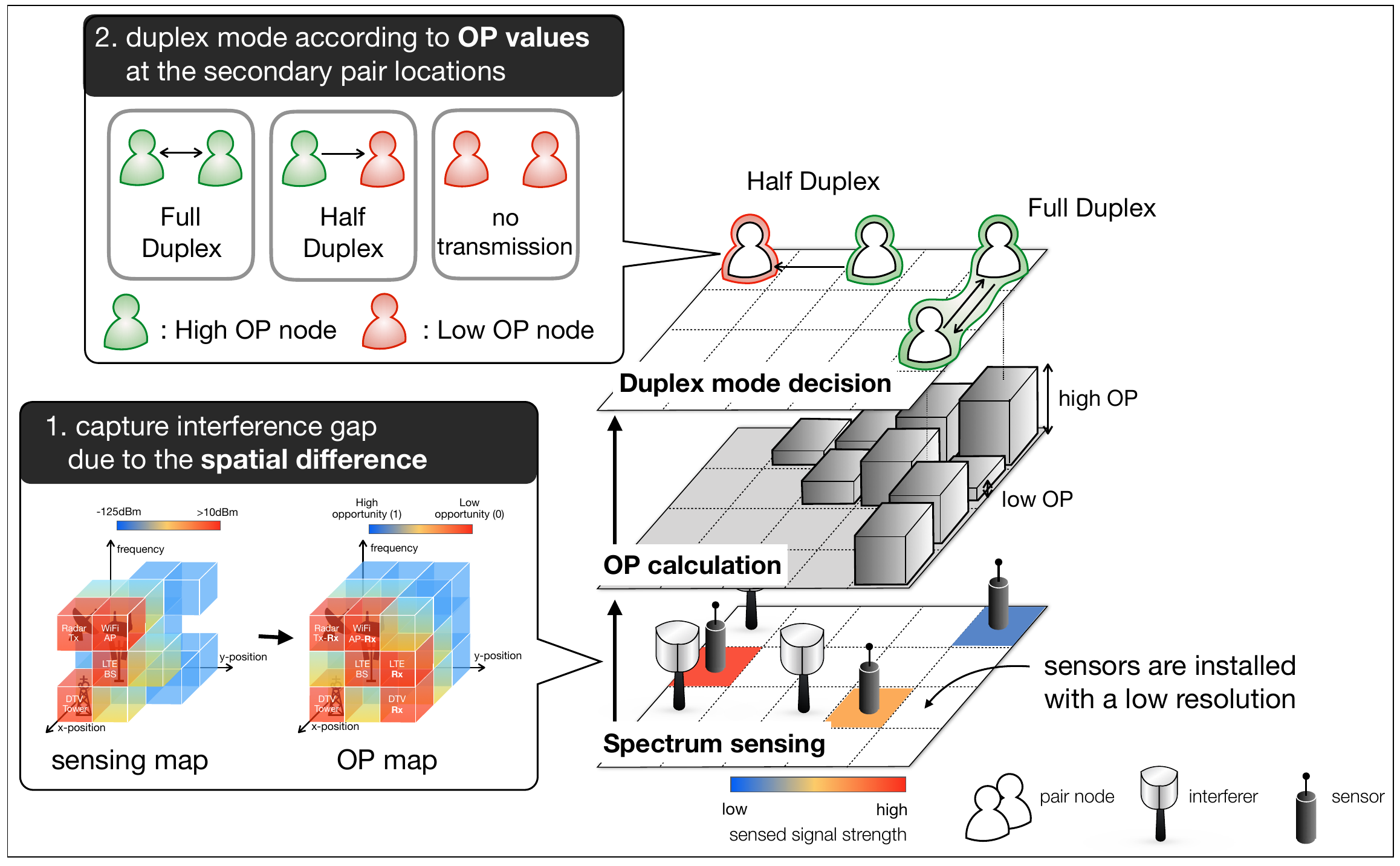}}
\caption{OP based hybrid HD/FD communications. \label{Fig:totalsystem}} 
\end{figure*}

\subsection{OP Detection}
This article proposes the OP detection algorithm in which given the channel sensing result, the OP value at each node's location is calculated based on the predicted level of interference, i.e., $\text{OP} :=\mathsf{P} \(\SIR>\theta | I\),$ where $\theta$ and $I$ represent the access threshold and the measured interference level at the nearest sensor, respectively.
Note that the access threshold $\theta$ implies the transmission requirement, so the OP represents the predicted probability of the transmission success.

To represent the OP, we focus on the spatial correlation of interference between the node and its nearest spectrum sensor.
The amount of correlation is highly contingent on the spatial difference between the node and the sensor; this changes in a topological environment, as with interferer locations.
Specifically, when the node and the sensor are co-located, it is obvious that they meet the same level of interference.
As the distance between them lengthens, the similarity decreases and finally they become independent.
In our previous study \cite{dyspan}, we figured out, with the aid of stochastic geometry (SG), this spatial correlation in the form of the conditional probability.

It is worth noting that this probabilistic approach plays a role in transforming the real interference value to the level of the OP in a stochastic manner. 
The measured interference level at the sensor itself cannot provide an accurate prediction of the interference level at the node because it depends on various parameters, such as the density of TXs, and transmit power \textit{et al}.
For instance, even though the measured interference at the sensor seems to be relatively high, there can be more OPs around the node especially when the density of TXs is low. 
SG is able to reflect these kinds of parameters within one metric in a probabilistic form, which thus harmonizes with the operation of the proposed OP based  spectrum sharing system.


\subsection{Fundamentals of MAC for Spectrum Sharing Networks}

In this subsection, we examine some criteria for designing a MAC algorithm. An intuitive way to maximize spectral efficiency is to select pairs of the nodes optimally. Such a centralized manner requires, however, a coordinator to exactly know the locations of nodes and real-time channel qualities of all communication links. Because it is hard to be realized in practice, we consider a distributed manner instead of the centralized operation to design a practical MAC algorithm. 

In sharing networks, the nodes should not transmit with their maximum power in every time slot for coexistence and harmony with other networks (e.g., primary protection in cognitive radio (CR) networks). 
We can infer the level of interference using the OP value in our framework, and the presumption of interference level helps us to judge the success probability of communication of the node. Hence, we adjust the transmission probability using the OP value in our MAC algorithm. In the following paragraph, we discover why we control the transmission probability instead of the transmission power. 
Fig.~\ref{Fig:ProbvsPower}a shows two shapes of the increasing function with solid lines; log-shaped (concave) function and linear function. When the product of node density and average OP of nodes is smaller than a certain threshold ($\beta$)\footnote{Due to space limitation, please see \cite{JMtwc} for details of increasing functions and the derivation of $\beta$.} 
the concave function is superior because the aggressive attempts lead to the higher throughput in spite of the increased interference. 
The linear function is better when the product is greater than $\beta$, because excessive transmissions may ruin the successful transmissions due to the immoderate interference. In the same figure, there are dotted lines that signify the threshold-based deterministic decision, and the threshold can be set by minimizing the difference from the solid line, i.e., probabilistic (random) transmission. As noted above, however, a deterministic decision of the transmission may cause severe interference and collisions due to the spatiotemporal correlated OP values of the nodes.
In this article, we use a linear function for the transmission probability by assuming spectrum-shared massive IoT networks that have high node density and moderate SINR threshold and link qualities.

Let us focus on two transmission pairs in the network as shown in Fig.~\ref{Fig:system}, of which the OP of a receiver is 0.8 and that of the other is 0.5. Figs.~\ref{Fig:ProbvsPower}c and \ref{Fig:ProbvsPower}d show the SINR regions of two receivers for the cases where the mutual interference is small and large, respectively. Each figure has three distinguishable points that denote SINR averages of each access mode described in Fig.~\ref{Fig:ProbvsPower}b. In Fig.~\ref{Fig:ProbvsPower}c, maximal power transmission shows the highest sum of SINR among three cases. Though reduced power transmission and random access show slightly smaller values than that, they produce less interference because the nodes spend a smaller value of \emph{(transmission time)} $\times$ \emph{(transmission power)}. We see that random access increases the average SINR of the node with a smaller OP (\emph{y}-axis) compared to the power reduction scheme. It comes from the scheduling effect of the probabilistic access.  
From the low OP node's viewpoint, the node has an opportunity of communication without interference from the high OP node (\emph{x}-axis) in the random access. That means the low OP node can communicate with the smallest interference for a given external interference. In the reduced power transmission, however, the interference from the high OP node always exists, and there is no possibility of SINR improvement. The mutual interference mitigation is much more of a benefit for nodes with low OP. In Fig. \ref{Fig:ProbvsPower}d, the random access shows higher sum-throughput than the maximal power transmission as the mutual interference becomes large. 
From this, we can infer the random access is more efficient in interference management and it ensures higher fairness than the reduced power transmission. 
Based on what we have so far, we use the random access as the basis of our MAC. 

For the ways of random access, we can consider CSMA or ALOHA that is a typical distributed MAC scheme, and there is a difference that CSMA's contention operation has dependency in time axis whereas ALOHA is independent. In snapshots, however, we can model active transmitters by stochastically thinning for both of the operations \cite{JHwang16}, so we analyze the ALOHA-like scheme for the analytical convenience. The result can be directly applied to the ALOHA scheme, but it can also be applied to the CSMA scheme by adjusting the contention window size so that the average transmission probability is the same. 

Now, we check the decision rule for the transmission probability when FD is available. Let us consider a single pair in the network. Both transmit to each other simultaneously in the FD case, whereas only one node transmits in a single time slot in the HD case. By comparing the throughput of both cases, we know that the throughput gain of FD becomes larger as the interference decreases. However, HD is better when the interference is small, because the self-interference (SI) caused by the FD communication ruins the high quality of the communication link in this case. 
From this, we can understand that the communication pair with high OP nodes should have high opportunity for FD transmission when the interference is higher than a certain level (or the performance of SI cancellation is high enough). 

From the fundamentals that we examined, we design our MAC to set the transmission probability of the nodes in proportion to their own OP value whether FD is possible or not.

Fig.~\ref{Fig:totalsystem} illustrates the operation of our OP based hybrid HD/FD communications. 
	The server first calculates the OP value based on the sensed results from sensors, and then decides the duplex mode of each pair.
	For instance, when the pair nodes have $0.8$ and $0.6$ as their OP values respectively, the pair utilizes the FD with a probability of $0.48$ since their transmission probabilities are equal to the OP values.

\begin{figure*}[t!]
	\centerline{\resizebox{2.0\columnwidth}{!}{\includegraphics{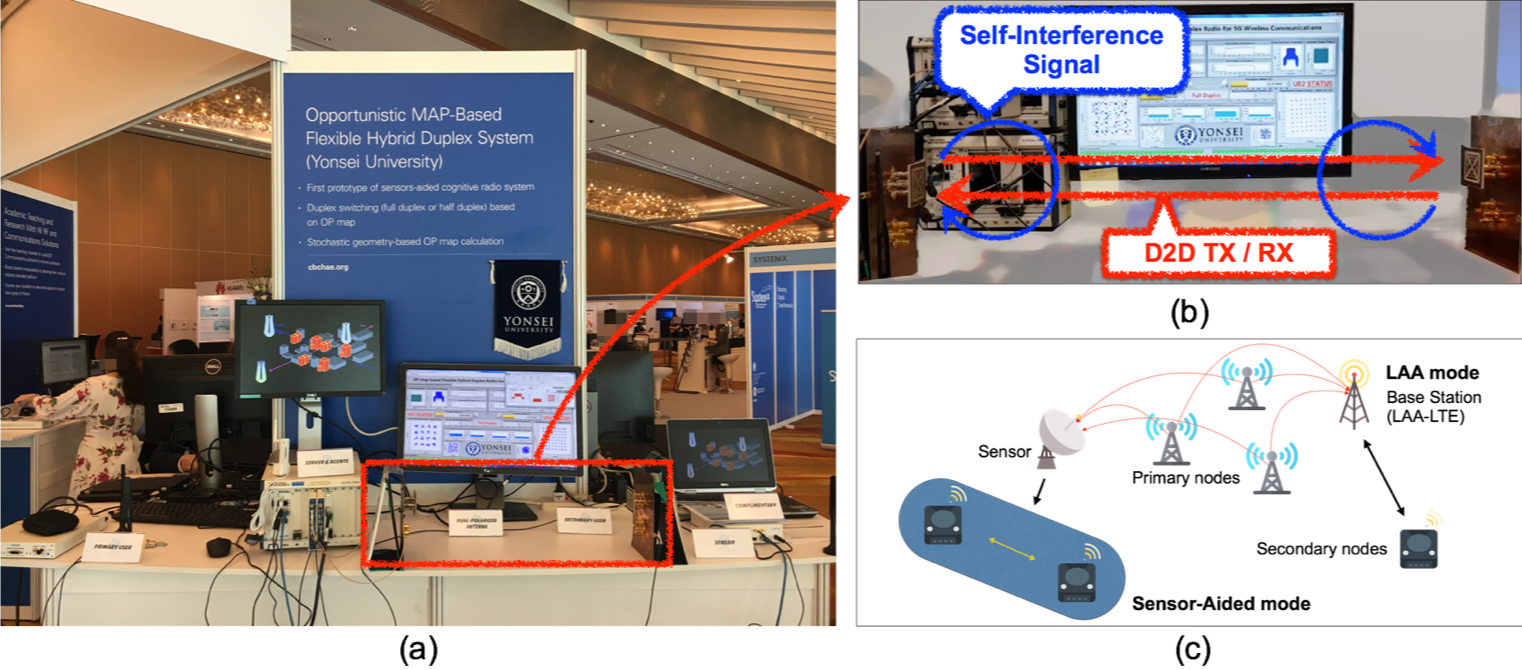}}}
	\caption{a) Real-time demonstration at IEEE Globecom, Singapore in Dec. 2017; b)  a basic concept of full duplex (self-interference signal and device-to-device signals); c) two representative modes in the spectrum sharing network based on the spectrum sensing, which are Sensor-Aided mode and LAA mode.}	
	\label{Fig:FD}
\end{figure*}

\subsection{Practical Methods for Spectrum Sharing Network with Full Duplex}


To implement FD in practice, it is desired to cancel the SI, which is the transmitted signal's echo to its own receiver, to the noise floor as shown in Fig. \ref{Fig:FD}b. Accordingly, researchers often focus on cancelling out the maximum amount of the SI. Based on advanced SI cancellation techniques, several studies have been underway to verify the concept in practice. Researchers have proved that the actual results differ significantly from what the theories propose \cite{Sachin, MKFD, MKFD2}.



With the advent of FD feasibility, Listen-and-Talk (LAT) strategy, which transmits and senses spectrum usage simultaneously, was introduced \cite{LAT}. This beneficial method, however, still has several problems such as additional energy/cost consumptions of the secondary nodes and cognitive capability for wideband sensing in a user terminal. This hinders the adoption of the LAT tactic in device-to-device (D2D) scenarios. The sensing performance deteriorates due to the residual transmitting signal.

To overcome these obstacles, we propose an algorithm that could employ not only base station but also D2D scenarios by having no power consumption.
Further, if only a few fundamental issues of FD such as low power and miniaturization are resolved, it has more advantages in being applied to the IoT environments. It could also achieve additional throughput gain with a flexible hybrid duplex system. 
The algorithm operates in the sensor-aided mode as shown in Fig. \ref{Fig:FD}c (while LAT represents LAA mode in the figure). Thus, it can work in D2D and gain such advantages like exploiting cooperative sensing technique in dense areas \cite{sensor}. In addition, eliminating the sensing inaccuracy with the exact target, the most fatal bane of this mode, is guaranteed using OP with high reliability.

In the following section, we illustrate the entire prototype setup with our proposed algorithm, which is one sensor-based opportunistic algorithm with random MAC decisions. 
The main contribution of our experiment is its validation of the feasibility (see Fig. \ref{Fig:FD}a\footnote{Full demonstration video is available at http://www.cbchae.org/.}) and providing the characteristics of FD in a network sharing system.

Over the past decade, research efforts in this area have shown great potential, but are mostly limited to academia, due to the difficulty of accurate spectrum/spatial sensing. To deal with these constraints (e.g., multipath fading and shadowing), many consider cooperative sensing to be a good solution. However, it cannot entirely address them with the most advanced technology, since delay and cost occur through decision making and data exchanging. On top of that, the accurate sensing is not yet sufficient to meet the needs of commercial deployments, due to the several additional limitations (e.g., inherent uncertainty of noise and vulnerability of the secondary node to the primary node emulation attack owing to the mandatory factor that the secondary node must guarantee the access of the primary node) \cite{dynamic}.
We propose one sensor-based algorithm that communicates if a certain level of probability is satisfied for a sufficient threshold, even if the primary performance is guaranteed to a lesser degree.


\begin{figure*}[t!]
	\centerline{\resizebox{2.0\columnwidth}{!}{\includegraphics{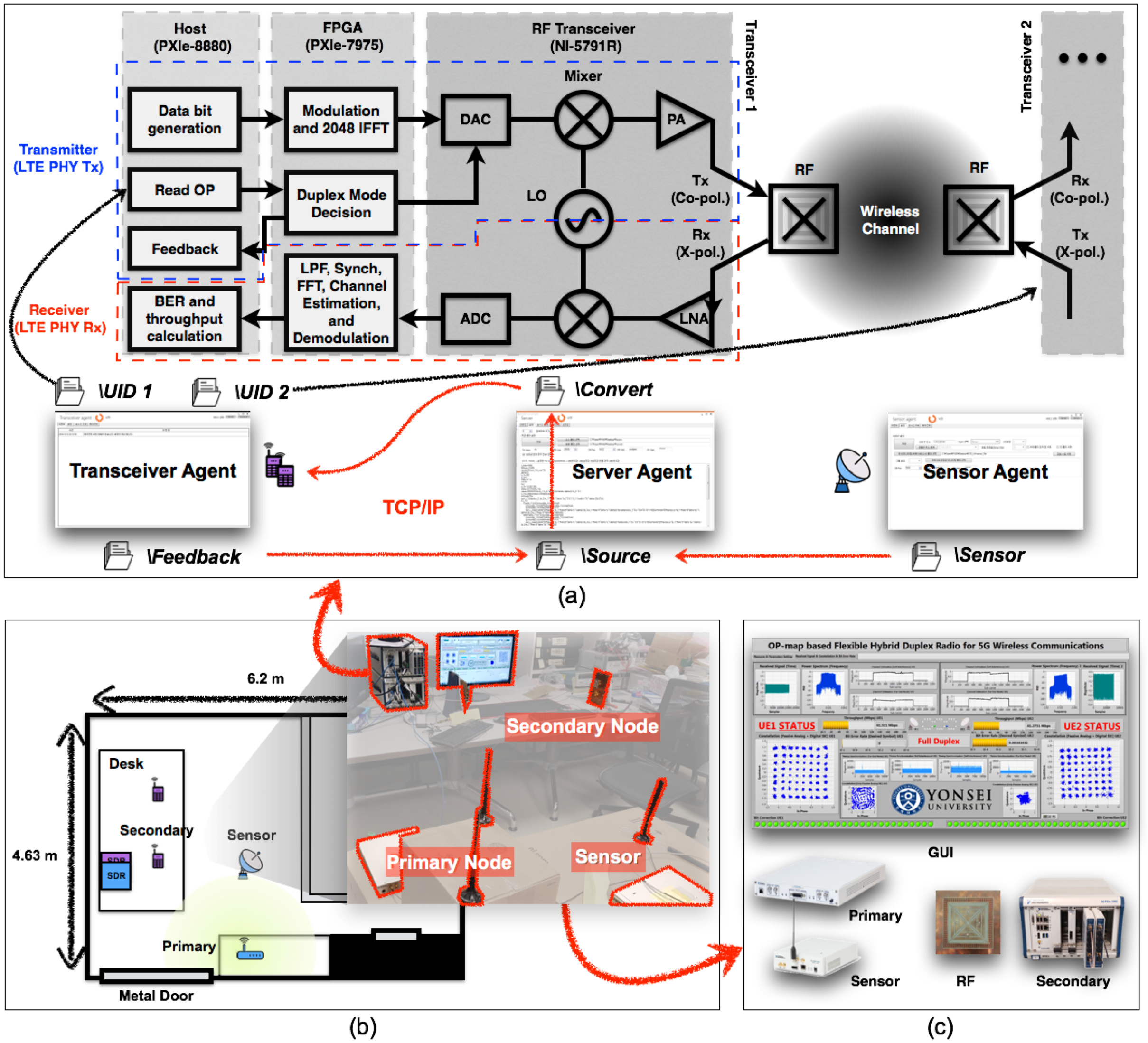}}}
	\caption{a) Block diagram of our proposed real-time testbed with data exchanging through the agents; b) a physical setup of our prototypes in an indoor testbed; c) a GUI of the SDR, hardware equipment of the primary node, the sensor, the dual-polarized flexible hybrid duplex front-end, and the transceiver respectively.}	
	\label{Fig:Design}
\end{figure*}

%
%
%
%
%

\section{Testbed Design}

\subsection{Hardware Architecture}

	The communication system, including all node terms such as the primary node, the secondary node, and the sensor is implemented using LabVIEW system design software and an FPGA based PXIe SDR platform as shown in Figs. \ref{Fig:Design}b and \ref{Fig:Design}c \cite{NI}. Most notably, the FD nodes that play a role as the secondary nodes consist of the following four main components:
	\begin{enumerate}
		\item A dual-polarized RF front-end for passive analog SI cancellation using a high cross-polarization discrimination (XPD) without any power consumption \cite{dualpole}.
		\item PXIe-8880 real-time (RT) controller for controlling the application and performing basic baseband processes that supports octa-core processor.
		\item NI 5791 RF transceiver module for providing dual 130 MS/s capability with 100 MHz bandwidth and 14/16-bit accuracy (ADC/DAC respectively).
		\item PXIe-7975 field-programmable gate array (FPGA) module, coupled to the 5791 RF, for high-throughput baseband processing with a Kintex-7.
	\end{enumerate}
	Note that, 1), 3), and 4) items are used for each FD node.
	
	The primary node is made up of USRP RIO 2953R with FPGA module, and the same operating controller, based on a selected subset of the 3GPP LTE Release 10. The deployed sensor that operates in 20 MS/s, consists of USRP 2922 with individual IP address.
	
	Moreover, all these modules are equipped in NI PXIe-1082 chassis which plays a key role in data aggregation with both FPGA processor sides and an RT controller for real-time signal processing. Above all else, it also supplies synchronizing frequency offset (CFO) to enable data processing precisely.
	
	As shown in Fig. 4a, from the sensor to the transceiver, via the server, transmitting data such as OP and sensing DB are transmitted by using the Transmission Control Protocol/Internet Protocol (TCP/IP), with our several implemented agents.	
	
	\subsection{OP Computation and Update Procedure with Transceiver}
	\label{method0}	
	
	With our designed blocks (see Fig. \ref{Fig:Design}a), we describe our real-time proposed testbed with the sequence order. Note that the sensor and the transceiver are in a fixed location, and every single datum exchanged through agents, is sent over the TCP/IP link. Feedback data also refers to the duplex decision information for each secondary node.
	\begin{enumerate}[1.]
		\item While the primary is off, calculate the distances between the sensor and the secondary nodes by measuring power for each of the four modes (FD, two kinds of HD, and silence).
		
		\item The sensor measures the data every 1 ms (sensing range (20 MHz) $\times $ acquisition time (1 ms)/IQ rate (20 MS/s)).
		
		\item In parallel with Step 2, the transceiver initially sends the feedback to the server with the silence mode.
		\item Servers compute the OP for each secondary node with the latest data in the sensing DB and the feedback (since the sensing DB changes much more frequently than the feedback, the operation cycle of computing OP is 1 ms).
		\item Each secondary node considers the computed OP to determine the duplex mode. 
		\item Each secondary node sends the feedback to the server.
		\item Proceed iteratively from Step 4 to Step 6.
	\end{enumerate}
	
	More specifically, the secondary nodes performing FD communications have reference signals and several specifications that are set to LTE standard to operate in the duplex mode. For simplicity, we assume D2D link structure as the frame structure of the LTE downlink which are based on 2048 orthogonal frequency division multiplexing (OFDM) symbols with extended cyclic prefix (CP, 512 length). Therefore 1 half frame (5 ms) comprises 10 slots which are made up of 6 symbols each. With our proposed algorithm, each node decides whether to transmit data using a duplex mode decision block (see Fig. \ref{Fig:Design}a) in every single half frame (5 ms). If the duration between the two nodes is out of order, or it is lower than 5 ms, data transmission may not be performed properly. This is because it is imperative to maintain the difference between the SI and the desired signal’s over-the-air propagation delays within CP duration in the FD system \cite{MKFD2}, and each half frame includes exactly one primary synchronization signal (PSS). Note that synchronization signal that uses Zadoff-Chu sequences, placed in sixth OFDM symbol of every first slot of the half frame. Several blocks using Xilinx IP are also implemented in an FPGA domain; these blocks are a low-pass filter (LPF), a fast Fourier transform, and a channel estimation block that exploits the pattern of cell-specific reference signals. Zero forcing based channel equalizer block is also exploited in the FPGA domain.
	
	%
	%
	%
	
	\subsection{Test Scenario}
	\label{method1}
	
\begin{figure*}[t!]
		\centerline{\resizebox{2.0\columnwidth}{!}{\includegraphics{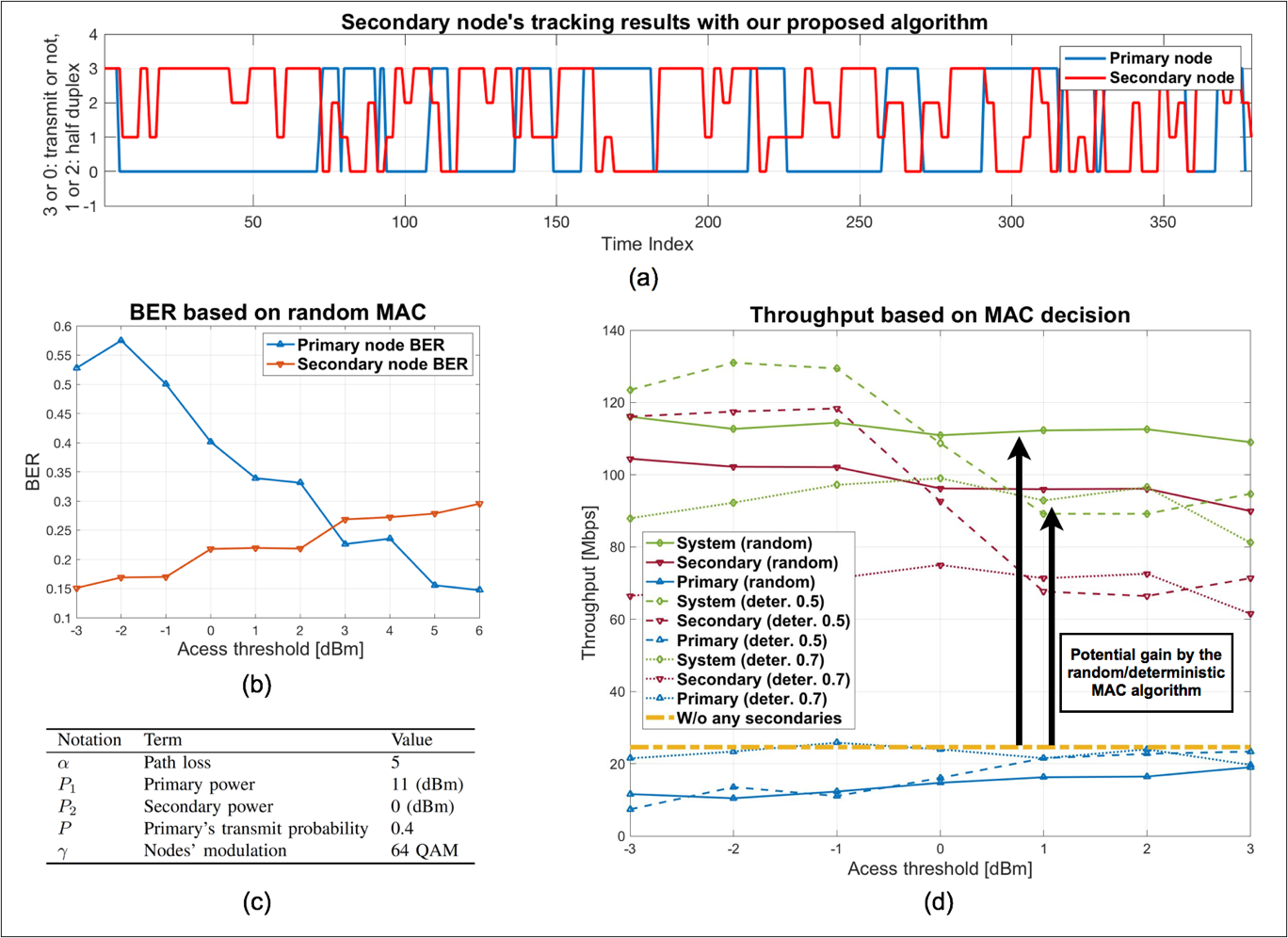}}}
		\caption{a) The secondary node's tracking results in our proposed algorithm; b) BER of the secondary node and the primary node with the increment of the access threshold; c) range of parameters used in the analysis; d) the system throughput, the secondary node's throughput, and the primary node's throughput with the increment of the access threshold.}	
		\label{Fig:results}
	\end{figure*}

	We present the measurement campaign scenario that was conducted in Veritas Hall Building C, 332 at Yonsei University as shown in Fig. \ref{Fig:Design}b (4.63 m $\times$ 6.2 m). Through parameters as illustrated in Fig. \ref{Fig:results}c, with the fixed location of the secondary pair nodes and the sensor, we placed the primary node in unknown spots and measured for enough measurement time to collect the data more accurately. 
	
	First, we measured the interference level at the sensor for the full duplex mode, the half duplex mode, and the silence mode. 
	Then we measured the system throughput of primary nodes and secondary nodes based on bit-error-rate (BER) performance in our testbed. 
	The throughput is calculated with the consideration of \emph{overhead factors}, which are the reference symbols, the PSS, and the CP.

	\section{{\fontsize{11}{14}\selectfont Experimental Evaluation}}
	\label{Sec.evaluation}
	
	In this section, we provide experimental results of our OP based flexible hybrid duplex system in an indoor testbed. From our experimental results, we verified and characterized the link-level performance of our proposed algorithm with the several conditions. We measured the time delay of each procedure before starting the experiment. The measured time delay of TCP/IP per link was about 1 ms, and took about 2.5 ms, even though the distance between Seoul (where the server is) and Incheon (where the testbed is located) was taken into account. The time delays of the feedback and the sensor were 5 ms and 1 ms, respectively. Thus, as in the situation briefly described below, in the worst-case scenario, the total time delay amounts to 5 + 2.5 + 1 + 2.5 + 4 = 15 (ms), except for the OP computation time (using a server in the same region, 5 + 1 + 1 +1 + 2 = 10 (ms)). In Step 1, with the primary condition changed, the time delay of the updated new feedback value can be up to 5 ms. Then, after the feedback is sent to the server in step 2, the server starts the OP calculation for up to 1 ms (It is the same as the sensor update period). The computed OP is sent to the transceiver in Step 4, and the arriving OP is considered in duplex decision block for 4 ms (the feedback time is a multiple of 5 ms). Despite the assumption of the worst-case, and even with all these time delays, the OP calculation time of about 132 ms using Matlab dominates over the proposed testbed system. Note that the system latency is limited by the TCP/IP, and Matlab calculations. This can be further optimized by implementing all these procedures in an FPGA chip and applying the low bound of OP calculation (17 ms) \cite{dyspan}.
	
	Taking all the above into consideration, the transceiver is set to 3 seconds to update reading OP period and the primary node is set to every 11 seconds to update a transfer decision with a probability of 0.4. Though it incurs more error rate, we investigated the characteristics of our proposed algorithm under the several conditions. The throughput with the access threshold in the random MAC decision is investigated in Section A, while the throughput with the deterministic MAC decision is investigated in Section B.
	
	\subsection{Impact of the Access Threshold}
	\label{method4}
	
	As can be seen in Fig. \ref{Fig:results}a, the $y$-axis represents the current mode decision of each node, and we can verify that the tracking method works well. Also, with the increment of the access threshold, the system throughput and the secondary node's throughput are slightly reduced, while the primary is increased (see Fig. \ref{Fig:results}d, the solid lines). Note that the primary node transmits data with a probability of 0.4 and it incurs the throughput degradation of the primary node compared to the secondary node, while the secondary throughput includes two paired nodes’ throughput which could operate in the FD mode. The nodes' BER is also shown in Fig. \ref{Fig:results}b, and we can verify the primary node's BER decreases while the secondary increases. We only considered 0.4 probability of transmission in the BER graph of the primary node, and confirmed that the primary can be protected to some extent above the sufficient access threshold.
	
	\subsection{Deterministic MAC Analysis}
	\label{method5}
	
	With the result from Fig. \ref{Fig:results}d, the dashed lines were performed by the deterministic value of 0.5, which the node decides the transmission depending on whether the OP is greater than 0.5. The secondary nodes began to stop transmitting signals at a low OP (the primary's transmission), in access threshold of 0, which represents a high chance of transmission of the primary node. Therefore, it degrades the system and the secondary nodes' throughput, while increases the primary node's throughput. Thus, the primary node's throughput increases higher than the random MAC decision while the secondary decreases, because the deterministic MAC does not allow the transceivers to transmit signals. Also, the dotted lines representing the deterministic value of 0.7 indicates that the secondary nodes do not transmit signals if there is a transmission of the primary node in the access threshold from -3 to 2. From the access threshold~3, the deterministic MAC stops the secondary nodes from transmitting signals more frequently regardless of whether the primary node is transmitting a signal or not. 
	
	In summary, compared to the dashed/dotted line representing the throughput of the primary node without any secondary nodes, the primary node is guaranteed to be an ideal value at the high access threshold or the proper deterministic value, but the random MAC should consider some performance degradation. From the system level perspective, however, the random MAC ensures stable high performances even for a single pair analysis, whereas the deterministic MAC exhibits a relatively much lower performance.
	Moreover, since the access threshold is the transmission requirement, it is not an easy task to iteratively control the access threshold based on the results. Thus, the random MAC that is less affected by the access threshold and generally provides good results is better than the deterministic MAC, which occasionally produces good results depending on the access threshold.

	\subsection{Challenge: Multi-device Analysis}
	We jointly implemented our proposed algorithm with a single pair secondary node. It is obvious that there are wide discrepancies between multiple devices and a single device analysis, specifically random MAC vs. deterministic MAC as analyzed above. In addition, it will be interesting to explore cooperative sensing using multi-sensors as well as evaluating the performance of multi-nodes in a wider space.

	\section{Conclusion}
		This article proposed a spectrum sharing MAC to support massive IoT networks in 5G. It was based on an OP framework that represents the spectrum usage level in a probabilistic manner. We first validated the feasibility of random MAC in which each node accesses the spectrum with a probability that proportionally increases with its OP value. Utilizing the OP based random MAC, we also proposed hybrid HD/FD communications where the duplex mode of each pair is dynamically determined based on the OP value. With the link-level evaluation based on a software-defined radio testbed, we confirmed that the proposed algorithm showed great potential in several conditions. We expect our study to provide in-depth insights into PHY/MAC design of dynamic spectrum access for 5G.
		
	\section*{Acknowledgment}
	This work was supported by Institute for Information and communications Technology Promotion (IITP) grant funded by the Korea Government (MSIT) (No. 2015-0-00294, Spectrum Sensing and Future Radio Communication Platforms).


\end{document}